\documentclass{article}

\usepackage{arxiv}

\usepackage[utf8]{inputenc} 
\usepackage[T1]{fontenc}    
\usepackage{hyperref}       
\usepackage{url}          
\usepackage{natbib}
\usepackage{booktabs}      
\usepackage{amsfonts}   
\usepackage{nicefrac}         
\usepackage{lipsum}		
\usepackage{graphicx}
\usepackage{doi}

\usepackage{setspace}
\usepackage{float}
\usepackage[utf8]{inputenc}
\usepackage{lscape}
\usepackage{adjustbox}
\usepackage{tabularx}
\usepackage{array}
\usepackage{tablefootnote}
\usepackage{amsmath}
\usepackage{caption}
\usepackage{threeparttable} 
\usepackage{multirow}
\usepackage{lipsum}
\usepackage{makecell}  
\usepackage[singlelinecheck=false]{caption}
\newcolumntype{C}{>{\centering\arraybackslash}X}

\usepackage{graphicx,verbatim}

\title{++nnU-Net: Scaling nnU-Net with Prefix-Based Data Augmentation}

\author{
    \parbox{\linewidth}{\centering
        \textbf{Ana Sofia Santos}$^{1,2}$, \textbf{André Ferreira}$^{1,2}$, \textbf{Gijs Luijten}$^{2}$, \textbf{Naida Solak}$^{2}$, \\[0.3em]
        \textbf{Lisle Faray de Paiva}$^{2}$, \textbf{Behrus Hinrichs-Puladi}$^{3}$, \textbf{Jens Kleesiek}$^{2,4}$, \textbf{Jan Egger}$^{2,4}$, \textbf{Victor Alves}$^{1}$ \\[0.8em]
        \mdseries
        $^{1}$Center Algoritmi / LASI, University of Minho, Braga, Portugal \\[0.2em]
        $^{2}$Institute for Artificial Intelligence in Medicine, University Medicine Essen, Essen, Germany \\[0.2em]
        $^{3}$Institute of Medical Informatics / Dept. of Oral and Maxillofacial Surgery, University Hospital RWTH Aachen, Germany \\[0.2em]
        $^{4}$Faculty of Computer Science, University of Duisburg-Essen, Essen, Germany
    }
}

\date{}

\hypersetup{
pdftitle={A template for the arxiv style},
pdfsubject={q-bio.NC, q-bio.QM},
pdfauthor={David S.~Hippocampus, Elias D.~Striatum},
pdfkeywords={First keyword, Second keyword, More},
}

\begin{document}
\maketitle

\begin{abstract}
The nnU-Net has demonstrated continuous success in medical segmentation tasks, which heavily rely on the availability and diversity of annotated biomedical data. However, assembling medical imaging cohorts remains challenging due to numerous factors such as privacy regulations and annotation costs. As a result, data augmentation plays a crucial role in increasing data availability while maintaining anatomical feasibility. Hence, we propose the ++nnU-Net, a novel data augmentation module based on image registration that operates prior to preprocessing and training take place. Our framework was evaluated across five different 2D datasets. In this workflow, image data go through a two-stage registration process, generating new warped images. The transformations are then applied to the respective segmentation. In addition, the pipeline computes available disk space, generates supplementary binary synthetic masks and generates checkpoints. We demonstrate that the ++nnU-Net outperforms the nnU-Net baseline, yielding improvements in Dice Similarity Coefficient scores. In the most prominent cases, we observe performance gains of approximately 22\%. These findings highlight the effectiveness of registration-based data augmentation, particularly for 2D medical imaging datasets and suggest that the ++nnU-Net provides a practical and scalable approach for enhancing segmentation performance in data-limited settings. The source code for the ++nnU-Net is available at: \url{https://github.com/sofia-adelie/plusplusnnunet.git}

\keywords{nnU-Net \and Data Augmentation \and Image Registration.}

\end{abstract}

\section{Introduction}

The continuous success of the nnU-Net \cite{isensee2021nnu}, has had a 
tremendous impact  in (medical) machine learning communities. In general, 
algorithms are primarily developed in a non-medical computer vision community 
and then transferred and adapted for medical tasks, such as the Fully 
Convolutional Network (FCNN) \cite{long2015fully}. In contrast, the U-Net 
\cite{ronneberger2015u} was originally proposed for biomedical image 
segmentation, being evaluated on microscopy images of cells, and has since 
been widely adopted by the broader computer vision community.
The nnU-Net is one of the rare cases where it was the contrary. It was 
primarily developed for medical scenarios and then widely adapted by the 
general computer vision community. The neural network simplifies the usage of 
the standard U-Net \cite{ronneberger2015u} and its outstanding performance in 
medical tasks has been repeatedly proven \cite{isensee2021nnu}. 
It streamlines the several processes such as preprocessing, training and 
inference. It was successfully evaluated in the Medical Segmentation Decathlon 
challenge, where it adapted to ten diverse datasets and achieved the highest 
Dice Coefficient scores in most cases, ultimately winning the competition with 
a significant margin \cite{decathlon2022}. Although the nnU-Net has shown remarkable success in multiple datasets, its performance, as in all deep learning models, is highly dependent on the quantity of the biomedical images collected \cite{lunardo2025}.  For instance, data augmentation has proven substantial impact in relation to the performance of deep learning models, with best performance relating to data augmentation for the heart, lung and breast \cite{garcea2023}.
By augmenting the available training data through various transformations, data augmentation addresses issues of limited sample sizes and dataset bias. This is particularly important in biomedical image segmentation tasks, where the acquisition of diverse and well-annotated datasets can be challenging \cite{garcea2023,goceri2023}. 

To address these challenges, we propose a data augmentation modular system that paired with the standard nnU-Net, optimizes performance compared to baseline. Hence, we introduce the ++nnU-Net, which incorporates a data augmentation module based on image registration. Because this type of data augmentation has been proved in various 3D settings, we restricted our work to 2D datasets.

\subsection{Related Work}
The implemented data augmentation pipeline followed the winning solution of the AutoImplant 2020 Challenge by Ellis et al. \cite{ellis2020deep}. They introduced an Advanced Normalization Tools (ANTs) based data augmentation technique that leverages non-rigid image registration to generate variations of CT skull data. The method pairs existing skull scans with their warps based on the \textit{antsRegistrationSyNQuick.sh} \cite{tutorialants} script from the ANTs \cite{serouj2025}. The process generates unique skull variations that significantly expand the size of the training dataset. The results showed a significant improvement in the performance of the authors' proposed model compared to baseline. By preserving crucial anatomical features while introducing variability in skull shape and size, this approach enhanced the model’s ability to generalize to diverse skull defects. 
This non-rigid registration framework has since been extended to large-scale 3D augmentation efforts in other domains. Notably, the same underlying approach was used to generate over 23,000 synthetic brain tumor images for glioma segmentation \cite{solak2025}, laying the foundation for the top-performing submission that won the BraTS 2023 challenge \cite{ferreira2024wonbrats2023adult}. 

In the same manner, Khajarian S. et al. \cite{serouj2025} introduced a data augmentation pipeline for liver tumor segmentation that combines non-rigid anatomical deformation with GAN-based CT synthesis. The process uses non-rigid registration to generate textured tumor variations masks out of two existing ones and then generating a CT volume for the corresponding new mask. The method proved to improve segmentation performance by approximately 3\%. Because the usage of data augmentation through image registration has been proved in various 3D settings, we restricted our work to 2D datasets.

\section{Methods}

This new system architecture accepts up to eight arguments, including the input directories containing images and segmentations, the desired of the final dataset after augmentation (i.e. original plus augmented data), the option to save a checkpoint documenting the registrations and the type of transformations during registration. Moreover, the user also has the option to preprocess their data to meet nnU-Net formatting requirements. If no transformation type is specified by the user, the default configuration is a two-stage registration comprising rigid and deformable Symmetric Normalization (SyN) transformations, the one used in this study during augmentation. Figure 1 displays how the data augmentation workflow operates, later described in the follow-up sections. 

\begin{figure}
    \centering
    \includegraphics[width=1\linewidth]{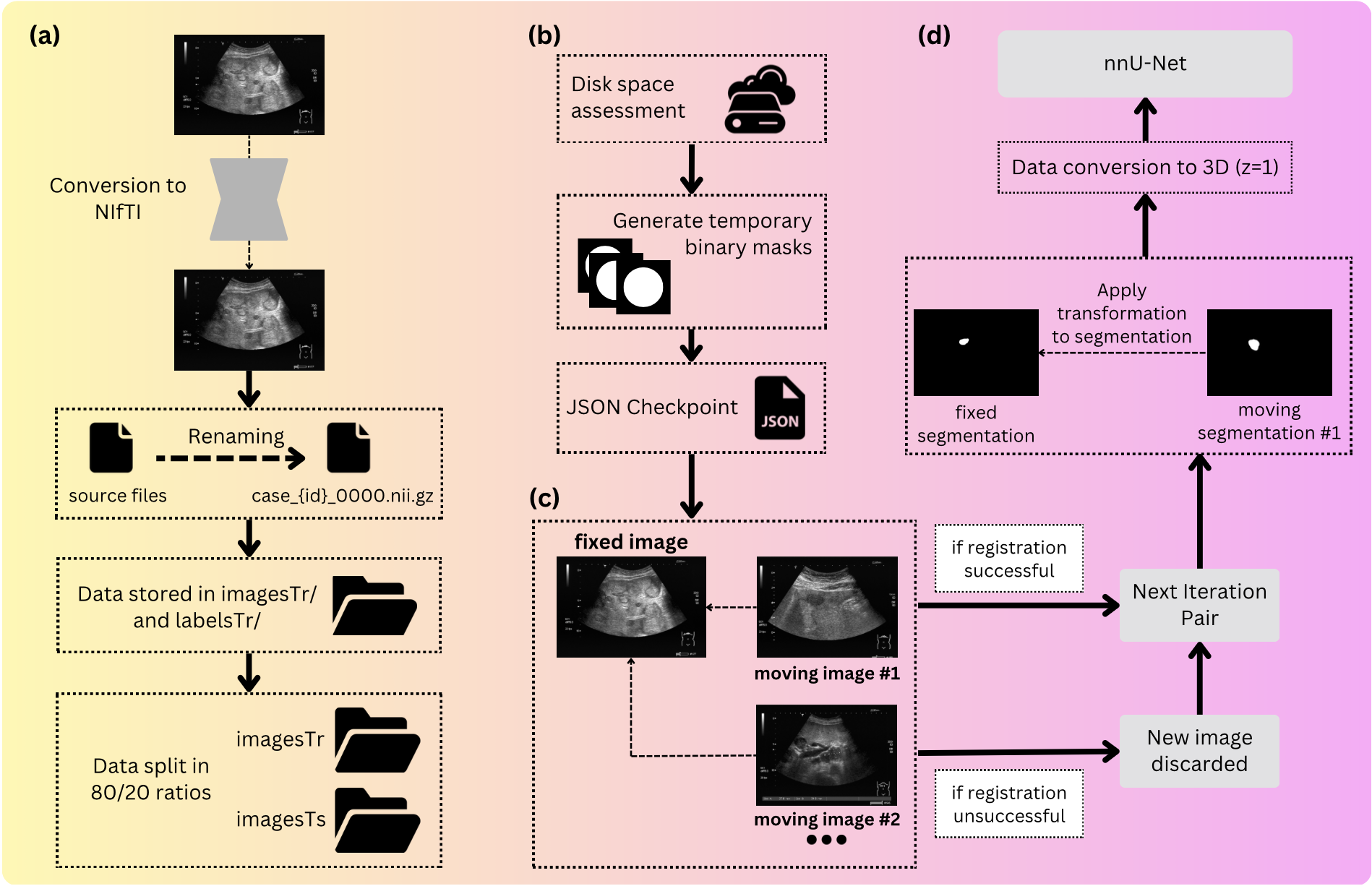}
    \caption{Overview of the augmentation pipeline: (a) data are converted to NIfTI format, renamed, stored in respective image and segmentation directories and split in training and test cohorts; (b) initial stages of the data augmentation process; (c) iteration process between images to be registered and respective transformation to generate segmentation masks; (d) final step of conversion to 3D format before training by nnU-Net.}
\end{figure}

\subsection{Data}

This study leveraged publicly available 2D datasets, with no private data, to ensure reproducibility. In total, we used five different datasets, as displayed in Table 1. In three of the datasets, the Pancreatic Liver Metastases Ultrasound (PLMUS) \cite{pancreaticUS}, the Breast Ultrasound Images Dataset (BUSI) \cite{busi2020}, and the Pelvic Fragment Segmentation on Synthetic X-ray Images (PENGWIN Task 2) \cite{pengwin}, not all original samples were used for the study. The Pancreatic Liver Metastases Ultrasound dataset is a collection of 105 ultrasound 
2D images and respective tumor masks, in which most images have a single ultrasound 
field of view. We then removed 6 images and respective segmentations from the 
original dataset since these contained two ultrasound acquisitions within a single 
image.

\begin{table}[ht]
\centering
\caption{Overview of the datasets and their properties used in the study (*).}
\vspace{10pt}
\label{tab:datasets}
\renewcommand{\arraystretch}{1.2}
\begin{threeparttable}
\setlength{\tabcolsep}{23pt}       
\begin{tabular}{l c c c c}
\toprule
\textbf{Dataset} & \textbf{Dimension} & \textbf{Modality} & \textbf{Size} & \textbf{Image Size} \\
\midrule
PLMUS \cite{pancreaticUS}  & 2D & US    & 99   & avg.\ 862 $\times$ 649 \\
BUSI  \cite{busi}          & 2D & US    & 647  & avg.\ 500 $\times$ 500 \\
ARCADE \cite{arcade}       & 2D & X-ray & 1297 & 512 $\times$ 512 \\
PENGWIN \cite{pengwin}     & 2D & X-ray & 250  & 512 $\times$ 512 \\
PDRD \cite{dental}         & 2D & X-ray & 598  & 2041 $\times$ 1024 \\
\bottomrule
\end{tabular}
\vspace{5pt}
\begin{tablenotes}
  \small
  \item[(*)] References indicate the dataset source used in this study, which may differ from the original publication introducing the dataset.
\end{tablenotes}
\end{threeparttable}
\end{table}

In the same manner, the BUSI dataset was originally utilized for both 
segmentation and classification tasks and is categorized into three classes: normal, 
benign and malignant. Because our primary goal concerns segmentation, we excluded 
the 133 images classified as ‘normal’ (i.e. ultrasound images of healthy tissue and 
without tumors). In addition, two other aspects were considered while handling this 
dataset: (1) the images categorized as ‘benign’ are significantly outnumbered 
compared to the ones classified as ‘malignant’; (2) some ‘benign’ (1 image) and 
‘malignant’ (16 images) data, had two or more segmentations within the same 
ultrasound image. Therefore, we removed the 17 images and respective segmentations. 
Ultimately, the BUSI dataset applied on this study had a resulting number of 647 
ultrasound images. 
The PENGWIN X-ray dataset, on the other hand, derives from the original PENGWIN CT dataset, having 50,000 original training X-ray images and respective segmentations. Because the dataset already comprises a large number of samples, it was inferred that the effects of data augmentation would not be as notable. Hence, we randomly selected 250 samples and respective segmentations from the original dataset. In addition, this is the only multi-class dataset we utilized in this study, comprising a total of four classes, including background. 
With regard to the ARCADE dataset \cite{arcade2024}, we utilized the train and test cohorts of the Stenosis subset and there was a need to do further work on the data, specifically on the segmentations. The segmentations were kept under polygon annotations in JSON files. Hence, we converted these coordinates into binary masks, which became the segmentation files for our study. Finally, the Panoramic Dental Radiography dataset (PDRD) \cite{dental} is a collection of panoramic X-rays with 32 classes. For our study we considered the dental segmentations as binary masks. After gathering the data, we organized and structured them according to the nnU-Net requirements. This process followed a similar workflow across all datasets. In general, all data were renamed, converted to NIfTI format, stored in respective images and segmentation directories and, ultimately, data were randomly split into train and test cohorts with 80/20 ratios. 

\subsection{Data Augmentation Tool} 

The proposed augmentation tool works as a prefix to the standard nnU-Net. Contrary to the augmentations performed during training, these occur before preprocessing and training take place. Moreover, although the \textit{antsRegistrationSyNQuick.sh} \cite{tutorialants} script is a well-established tool, it allows a sigle registration at a time, not entire directories. Our method contributes to a more scalable and automated approach, enabling consistent and efficient registration across whole datasets without the need for manual pairwise configuration. Figure 1 displays how the data augmentation workflow operates. 

Although the primary goal of this study is to provide a data augmentation tool adjacent to the standard nnU-Net, during data collection and preprocessing for our study we understood that datasets come in many formats and structures. Preparing each individual dataset for augmentation and further input on nnU-Net proved to be time consuming and unequal across all datasets. Therefore, adjacent to the data augmentation workflow, we introduced an option where users can structure their datasets according to the nnU-Net requirements. This process converts 2D datasets into NIfTI format, renames and stores them in the respective images and segmentations directories, ultimately splitting them in 80/20 ratio.

The script begins by estimating the user’s available disk space, providing an initial sense of how many augmentations can be feasibly generated. This estimate is based on the allocation of 80\% of the total disk capacity and dividing it by the average file size (in bytes) of the original dataset, ensuring a balance between the volume of augmentation and storage restrictions. The script proceeds to generate temporary binary masks that are stored in a temporary directory. During preliminary attempts with deformable SyN registration, recurrent distortions were observed along the edges of the augmented images, which were not representative of anatomically realistic variations. Therefore, these temporary masks allow the registration process to occur only within a certain area of the image. The binary mask is computed by creating a circular region in the center of the image with a radius equal to 75\% of half the width of the image. After that, the pixels inside the circular region are set to 1 and the pixels outside the region are set to 0. It is important to note that these masks are synthetic and do not correspond to anatomical regions.
Following the generation of temporary masks for the entire training set, a checkpoint JSON file is automatically created. This file serves three purposes: (1) it records the highest case identifier present in the original training set; (2) documents the fixed–moving image pairs selected during the augmentation process; (3) stores the index of the last fixed file that is augmented. Storing the highest case identifier is crucial, as the augmentation is intended to be performed exclusively among original, non-augmented images. After the checkpoint JSON is created, the augmentation process begins. 
The pipeline iterates over each image in the training directory, randomly selecting another image from the same set to serve as the moving image, which is then registered to the fixed scan. For each registration attempt, the corresponding temporary binary masks for the selected fixed and moving images are allocated. The registration script then enters a loop with up to five attempts to achieve a valid registration outcome. In order to determine whether an attempt is considered valid, two conditions must be satisfied: (1) the maximum orthogonal distance between the transformed image and the original image must not exceed half the minimum image dimension; (2) the transformed centers of images must remain within the original image boundaries. This prevents transformations that move the relevant anatomy out of the frame, even if the user chooses to use other forms of registration. If both conditions are met, the registration is considered valid and the resulting augmented image is saved with a filename that follows the original dataset’s naming convention, with incrementing IDs. Otherwise, the attempt is discarded, the resulting augmentations are deleted and a new attempt is initiated. Once the augmented images are generated, the same transformations are applied to the respective segmentation files. By the end of each augmentation, non-essential files such as transform matrices and displacement field files are deleted. 
As a final quality control step, the script verifies whether the transformed segmentation file contains labels. If the segmentation is empty, both the segmentation and its associated image are removed and the augmentation process proceeds to the next case. 

In case the user interrupts the augmentation process before reaching the ultimate number of augmentations initially determined, the user simply needs to provide an additional path to the initially stored checkpoint file. The script then returns the augmentation process from the index of the last fixed file. Finally, once augmentation is complete, all data are converted to 3D format with z=1. Because we trained on the first version of the nnU-Net, it processes only 3D data. 

\subsection{Training Protocol}

To evaluate the impact of the proposed augmentation method, a baseline comparison 
study was conducted. Training was performed on an NVIDIA RTX A6000 GPU, using PyTorch 2.4.1 with CUDA 12.1. Both baseline nnU-Net (v1) and ++nnU-Net were trained with 5-fold cross-validation over 110 epochs. For the ++nnU-Net, no fine-tuning or 
architectural changes were applied to nnU-Net. Additionally, parameters such as 
batch size and patch size were kept the same as the baseline. This rigorous control 
of training conditions ensures that any observed performance differences are 
attributable to the data augmentation method itself. 

\section{Results}
The performance of the proposed ++nnU-Net was quantitatively assessed against the baseline nnU-Net in five 
medical imaging datasets, previously explained. The results, including the Dice Similarity Coefficient (DSC) 
and Hausdorff Distance (HD95), are summarized in Table 2. The ++nnU-Net consistently achieved high DSC scores 
across all datasets, with the most pronounced gains observed in the PLMUS and ARCADE cohorts. Comparing the 
baseline DSC results, their performance on 10k training samples improved 22.08\% and 13.69\%, respectively. For 
instance, in the multi-class PENGWIN dataset, improvements were observed across all the classes, particularly 
in class (2), where DSC increased, approximately, 15.81\%. The remaining classes (1) and (3) had DSC score 
increments by roughly 0.41\% and 2.92\%, respectively. In addition, two other aspects are noted: (1) while 
there were substantial improvements in DSC score, a slight increase in HD95 was observed through all datasets; 
(2) regarding both X-ray datasets, above a certain threshold of augmented cases the DSC score starts decreasing.

\begin{table}[ht]
\centering
\caption{Overall performance comparison of Baseline and ++nnU-Net on DSC and HD95.}
\vspace{10pt}
\label{tab:datasets}
\renewcommand{\arraystretch}{1.2}
\begin{threeparttable}
\setlength{\tabcolsep}{18pt}   
\begin{tabular}{l cc cc cc}
\toprule
\multirow{2}{*}{\textbf{Dataset}} 
& \multicolumn{2}{c}{\textbf{Baseline}} 
& \multicolumn{2}{c}{\textbf{++nnU-Net (1k)}} 
& \multicolumn{2}{c}{\textbf{++nnU-Net (10k)}} \\
\cmidrule(lr){2-3} \cmidrule(lr){4-5} \cmidrule(lr){6-7}
& \textbf{DSC} & \textbf{HD95} 
& \textbf{DSC} & \textbf{HD95} 
& \textbf{DSC} & \textbf{HD95} \\
\midrule
PLMUS  & 0.4140 & 37.08 & 0.4921 & 39.66 & \textbf{0.5042} & \textbf{28.40} \\
BUSI   & 0.8037 & \textbf{33.74} & 0.8013 & 35.19 & \textbf{0.8055} & 34.29 \\
ARCADE & 0.3887 & 65.28 & (*) & (*) & \textbf{0.4419} & \textbf{65.25} \\
PDRD   & 0.9117 & 3.20 & \textbf{0.9409} & 3.22 & 0.9117 & \textbf{3.17} \\
\multirow{3}{*}{PENGWIN}
& 0.7499 & 24.10 & 0.7530 & \textbf{19.22} & \textbf{0.7620} & 21.17 \\
& 0.4403 & \textbf{122.86} & \textbf{0.5099} & 123.27 & 0.4745 & 144.23 \\
& 0.4586 & 125.77 & \textbf{0.4720} & \textbf{104.97} & 0.4261 & 113.57 \\
\bottomrule
\end{tabular}
\vspace{5pt}
\begin{tablenotes}
  \small
  \item[(*)] the results for ARCADE dataset on 1k augmentations are not presented due to the original dataset already containing nearly 1k training cases.
\end{tablenotes}
\end{threeparttable}
\end{table}

\section{Discussion}
The current study demonstrates a substantial improvement in DSC scores across all five datasets, when employing the data augmentation system based on rigid and deformable image registration, compared to the baseline nnU-Net configuration. This indicates that registration-based data augmentation applied to 2D datasets can effectively improve segmentation performance, rather than merely increasing data variability. These results align with prior studies reporting improvements in DSC following data augmentation through image registration. 
However, we also noted a common minimal increase in HD95, which is consistent with the increased shape variability introduced by deformable registration. This can improve generalization while occasionally amplifying local boundary deviations. The results further suggest that the impact of registration-based augmentation is more pronounced in smaller datasets, where limited data availability may otherwise constrain model generalization. Particularly, the datasets that initially had a lower baseline performance, benefited the most. For instance, comparing the baseline results of the PLMUS and the ARCADE datasets, with their performance on 10,000 training samples, the DSC scores improved 22.08\% and 13.69\%, respectively. These findings strongly support that this approach may be particularly effective in mitigating data scarcity and anatomical variability. Our findings also highlight the importance of optimizing the augmentation ratio. Notably, in both X-ray modality datasets, we observed that nnU-Net achieved the highest DSC score with a moderate number of augmentations, which decreased when training with 10,000 cases. Altogether, these findings suggest that modular data augmentation through image registration constitutes a robust and generalizable strategy for improving nnU-Net performance across heterogeneous 2D medical imaging datasets.

Although the results show great promise, the study was limited to only two image modalities, namely ultrasound and X-ray images. However, prior studies already indicate that this augmentation technique performs well on CT and MRI data. Future work could extend this augmentation system to additional imaging modalities and 3D datasets in order to assess its generalizability across a broader range of clinical scenarios.

\section{Conclusions}
In this paper, we propose a novel modular data augmentation through image registration system designed to improve the performance of nnU-Net. The proposed ++nnU-Net incorporates registration-based data augmentation prior to preprocessing, with the aim of improving segmentation performance, measured by the DSC and HD95 metrics. Most existing work employing similar methods, focus on 3D datasets and is typically limited to one anatomical structure. In contrast, our study extended these findings by demonstrating consistent performance gains across five distinct datasets comprising multiple anatomical representations, supporting the broader applicability of registration-based augmentation strategies. These results highlight the potential of registration-based data augmentation as an effective and generalizable strategy for boosting segmentation performance, particularly in data-limited settings.

%
%
%
\bibliographystyle{unsrt}   
\bibliography{mybibliography}

\begin{thebibliography}{10}

\bibitem{isensee2021nnu}
Fabian Isensee, Paul~F Jaeger, Simon~AA Kohl, Jens Petersen, and Klaus~H Maier-Hein.
\newblock nnu-net: a self-configuring method for deep learning-based biomedical image segmentation.
\newblock {\em Nature methods}, 18(2):203--211, 2021.

\bibitem{long2015fully}
Jonathan Long, Evan Shelhamer, and Trevor Darrell.
\newblock Fully convolutional networks for semantic segmentation.
\newblock In {\em Proceedings of the IEEE conference on computer vision and pattern recognition}, pages 3431--3440, 2015.

\bibitem{ronneberger2015u}
Olaf Ronneberger, Philipp Fischer, and Thomas Brox.
\newblock U-net: Convolutional networks for biomedical image segmentation.
\newblock In {\em Medical image computing and computer-assisted intervention--MICCAI 2015: 18th international conference, Munich, Germany, October 5-9, 2015, proceedings, part III 18}, pages 234--241. Springer, 2015.

\bibitem{decathlon2022}
Michela Antonelli.
\newblock The medical segmentation decathlon.
\newblock {\em Nature Communications}, 2022.

\bibitem{lunardo2025}
Febrio Lunardo.
\newblock How much data do you need? an analysis of pelvic multi-organ segmentation in a limited data context.
\newblock {\em Physical and Engineering Sciences in Medicine}, 2025.

\bibitem{garcea2023}
Fabio Garcea.
\newblock Data augmentation for medical imaging: A systematic literature review.
\newblock {\em Computers in Biology and Medicine}, 2023.

\bibitem{goceri2023}
Evgin Goceri.
\newblock Medical image data augmentation: techniques, comparisons and interpretations.
\newblock {\em Artificial Intelligence Review}, 56:12561–12605, 2023.

\bibitem{ellis2020deep}
David~G Ellis and Michele~R Aizenberg.
\newblock Deep learning using augmentation via registration: 1st place solution to the autoimplant 2020 challenge.
\newblock In {\em Towards the Automatization of Cranial Implant Design in Cranioplasty: First Challenge, AutoImplant 2020, Held in Conjunction with MICCAI 2020, Lima, Peru, October 8, 2020, Proceedings 1}, pages 47--55. Springer, 2020.

\bibitem{tutorialants}
Tutorial: Registering the brain/minds marmoset atlas to new data using ants.
\newblock \url{https://dataportal.brainminds.jp/ants-tutorial}.

\bibitem{serouj2025}
Serouj Khajarian.
\newblock Data augmentation for liver tumor segmentation using structure, texture, and contrast.
\newblock {\em Bildverarbeitung für die Medizin 2025}, 2025.

\bibitem{solak2025}
Naida Solak.
\newblock Gbm-reservoir: Brain tumor (glioblastoma multiforme) mri dataset collection with ground truth segmentation masks author links open overlay panel.
\newblock {\em Data in Brief}, 2025.

\bibitem{ferreira2024wonbrats2023adult}
André Ferreira, Naida Solak, Jianning Li, Philipp Dammann, Jens Kleesiek, Victor Alves, and Jan Egger.
\newblock Enhanced data augmentation using synthetic data for brain tumour segmentation, 2024.

\bibitem{pancreaticUS}
Jan Egger.
\newblock 100+ 2d us images and tumor segmentation masks.
\newblock \url{https://doi.org/10.13140/RG.2.2.36586.77761}.

\bibitem{busi2020}
Walid Al-Dhabyani.
\newblock Dataset of breast ultrasound images.
\newblock {\em Data in Brief}, 2020.

\bibitem{pengwin}
B.~Killeen, M.~Liu, P.-C. Ku, S.~Yudi, Y.~Liu, S.~Yibulayimu, G.~Zhu, X.~Wu, C.~Zhao, Y.~Wang, M.~Armand, and M.~Unberath.
\newblock Pengwin task 2: Pelvic fragment segmentation on synthetic x-ray images.
\newblock \url{https://doi.org/10.5281/zenodo.10913196}.

\bibitem{busi}
Walid Al-Dhabyani.
\newblock Breast ultrasound images dataset (busi).
\newblock \url{https://scholar.cu.edu.eg/?q=afahmy/pages/dataset}.

\bibitem{arcade}
Maxim popov.
\newblock Arcade: Automatic region-based coronary artery disease diagnostics using x-ray angiography images dataset.
\newblock \url{https://zenodo.org/records/10390295}.

\bibitem{dental}
Humans In~The Loop.
\newblock Teeth segmentation on dental x-ray images, 2023.

\bibitem{arcade2024}
M.~Popov, A.~Amanturdieva, and N.~Zhaksylyk.
\newblock Dataset for automatic region-based coronary artery disease diagnostics using x-ray angiography images.
\newblock {\em Scientific Data}, 2024.

\end{thebibliography}
\end{document}